\documentclass[showpacs,preprintnumbers,amsmath,amssymb,amsfonts]{revtex4}

\usepackage[usenames]{color}

\providecommand{\abs}[1]{\left| #1 \right|} %
\providecommand{\av}[1]{\left\langle #1 \right\rangle} %
\providecommand{\rbr}[1]{\left( #1 \right)} %
\providecommand{\mt}[1]{\mathrm{#1}} %
\providecommand{\mc}[1]{\mathcal{#1}}%
\def\ra{\rightarrow}

\def\lBG{\text{\tiny{BG}}}
\def\G{\text{\tiny{G}}}
\def\B{\text{\tiny{B}}}

\begin{document}
\draft
\date{\today}
\title{Comment on ``Critique of multinomial coefficient method for evaluating Tsallis and R\'enyi entropies" by A.S. Parvan}

\author{Thomas Oikonomou} \thanks{Corresponding Author}
\email{thoikonomou@chem.demokritos.gr}

\affiliation{Institute of Physical Chemistry, National Center for Scientific Research ``Demokritos", 15310 Athens, Greece}

\begin{abstract}
Parvan [arXiv:0911.0383v1]  \cite{Parvan2009} has recently presented some calculations in order to demonstrate the incorrectness of the results obtained from the generalized multinomial coefficients (GMC) presented in Ref. \cite{Oik2007}. According to Parvan, the aforementioned approach of studying maximum entropy probability distributions is erroneous. In this comment I demonstrate that Parvan's arguments do not hold true and that the obtained results from GMC do not present either mathematical or physical discrepancies.

\end{abstract}

  \narrowtext
\newpage \setcounter{page}{1}

\maketitle

The fundamental flaw Parvan believes to have discovered in Ref. \cite{Oik2007} is exactly the same misunderstanding, with which E. T. Jaynes had once been acquainted \cite{Jaynes1965}. The subject matter is the difference between the Boltzmann and Gibbs entropies. In order to shed light on this issue, let us define $W$ as the number of the allowed energy states that a particle can occupy within an ensemble, and $p_i$ are the respective probabilities (or simply called particle probabilities) for each one of these particles. Then, the Boltzmann entropy $S_\B$ reads
\begin{equation}
S_\B=N \sum_{i=1}^{W}p_i\ln(1/p_i),
\end{equation}
where $N$ is the size of the collection (number of particles). Denoting the total number of configurations associated with the energy states by $\Omega$ ($\Gamma$ in the notation of Ref. \cite{Parvan2009}), we then assign the occurrence probability $P_j$ to each configuration. This results in the following expression for the Gibbs entropy $S_\G$
\begin{equation}
S_\G=\sum_{j=1}^{\Omega_\G}P_j\ln(1/P_j).
\end{equation}
with $\Omega\equiv\Omega_\G$. The two expressions given above are not always equal to each other \cite{Jaynes1965}. In fact, their equality is satisfied \textit{only} when the particles are independent from one another \cite{Jaynes1965}. If this is the case, then the total number of configurations in the Gibbs ensemble equals to that of the Boltzmann entropy i.e., $\Omega_\G\equiv\Omega_\lBG$. This property uniquely characterizes the BG-ensemble, and clearly distinguishes it from the Bose-Einstein-Gibbs and Fermi-Dirac-Gibbs ensembles. Accordingly, one can use $S_\B$ or $S_\G$ interchangeably in BG-collection of configurations. From the aforementioned equality, one can directly verify the extensivity of the BG-entropy. It is also well known, that from the combinatorial point of view $S_\B$ can be derived on the basis of the multinomial coefficient
\begin{equation}\label{MC}
\mt{C}_{n_i}^{N}=\frac{N!}{\prod_{i=1}^{W}(N p_i)!},
\end{equation}
in the limit $N\ra\infty$. Then, for equal probabilities one obtains the BG-configuration function depending on $N$,  namely
\begin{equation}\label{BG-ConfFun}
\Omega_\lBG(N)=\lim_{N\ra\infty}\mt{C}_{n_i}^{N}=\exp\Big(N\ln(W)\Big)\,.
\end{equation}
In order to find the thermodynamic expression of $\Omega_\lBG$ however, one should replace $W$ with physical
quantities, since the latter is of purely combinatorial nature.

Taking above explanation into consideration, one may think to generalize Eq. \eqref{MC} in order to derive a generalized entropic form. Indeed, Suyari first studied such a possibility in Ref. \cite{Suyari2006} with Tsallis (T) entropy being the point of departure. His success was partial in the sense that one could clearly see the connection between his generalized multinomial coefficient (GMC) and T-entropy but its exact structure could not be derived for all values of the generalization parameter. Following Suyari's steps, I continued along the same direction but in a different way in Ref. \cite{Oik2007} so as to define two GMC's (GMC$_1$ and GMC$_2$) in order to derive (among others) (Boltzmann-)Tsallis and (Boltzmann-)R\'enyi entropic structure and study their maximum probability distributions. Since the aforementioned entropic structures concern the generalization of the BG-ensemble, whether ones uses Boltzmann generalization or Gibbs generalization does not matter. Even from this much explanation, it becomes evident that Parvan's doubt of considering the entropy definitions in Ref. \cite{Oik2007} as incorrect is misplaced.\\

Let us now recapitulate the results obtained in Ref. \cite{Oik2007}. Regarding Tsallis entropy, it was shown that the
MaxEnt distribution structure obtained from GMC$_1$ is in accordance with the one obtained from Jaynes Formalism (JF) with ordinary linear constraints, $1/\exp_q(\abs{x})$ with $q\in[0,1]$, while the MaxEnt distribution structure obtained from GMC$_2$ is in accordance with the ones obtained from JF with escort linear constraints, $1/\exp_{2-q}(\abs{x})$ with $q\in[1,\infty)$. The $q$-intervals stem from the requirement of the positivity of the
GMC's, since a negative number of states is senseless. Unfortunately, reading again Ref. [2] after Parvan's critique, I realized that, although the equations themselves demonstrate that in Tsallis case (and only in this case) the results from JF and GMC's coincide, in the text right after Eq. (61) in Ref. [2], it states that all the results obtained from GMC's are in contradiction with the ones obtained from JF. This is a misprint, which might be the source of confusion on part of Parvan. This sentence only refers to R\'enyi and Gaussian entropy as it becomes obvious in the context later, where it is demonstrated that the combination of the results obtained from both methods, JF and GMC, sheds light on the violation of concavity of Tsallis entropy in its Escort Distribution Representation for $q\in[0,1]$, also discussed in Ref. \cite{Abe-Sisto}.\\

According to the results in Ref. \cite{Oik2007} the configuration function, or following Parvan's notation, the total number of microstates in Tsallis and R\'enyi ensemble are of the following form
\begin{align}
\Omega_{\text{\tiny{T}}}&=\exp_q(x),\qquad \Omega_{\text{\tiny{R}}}=\exp(x)\,,
\end{align}
which are in agreement with the respective entropy extensivity \cite{OikTirn2009}. Parvan, arbitrarily and erroneously, chooses to adopt Eqs. (27)-(29) in his critique \cite{Parvan2009}, i.e., the same $\Omega$ for all three ensembles, BG-, T- and R-ensemble. It is then straightforward that replacing for example a configuration function of ordinary exponential type in the Tsallis entropy for equal probabilities, one violates the extensivity property, as in Eq. (28) in Ref. \cite{Parvan2009}.\\

Regarding R\'enyi entropy, it is far more obvious that since the averaging procedure in R\'enyi entropy is of  exponential type and not of linear type as in BG-entropy i.e.,
\begin{equation}
\frac{S_{\lBG}}{N}=\displaystyle\av{\tau_i}_{\mt{lin}}=\sum_{i=1}^{W}p_i\tau_i\,,\qquad
\frac{S_{\text{\tiny{R}}}}{N}=\displaystyle\av{\tau_i}_{\mt{exp}}=
                 \frac{1}{1-r}\ln\rbr{\sum_{i=1}^{W}p_ie^{(1-r)\tau_i}}
\end{equation}
where $\tau_i:=\ln(1/p_i)$, the entire R\'enyi ensemble is characterized by the aforementioned non-linear averaging procedure. In fact, historically, R\'enyi obtained his measure by conforming to the requirement of additivity. The preservation of additivity leaves one with only two choices if one would like to keep the definition of information gain undeformed. These two choices are linear averaging and exponential averaging procedures. The former paves the way to BG entropy, whereas the latter yields to R\'enyi measure. In this sense, what R\'enyi did is remarkable, since one cannot write a new undeformed generalized entropy measure. R\'enyi saw the second possibility and that is it. However, it was not his intention to write a generalized entropy measure valid for a generalized Thermostatistics. This is a different agenda then just generalizing BG measure. Thus, in a thermostatistical framework, whenever one applies JF, or equivalently Parvan's formalism \cite{ParvanBiro2009} on R\'enyi entropy, one should consider the proper constraints, as correctly noticed in Ref. \cite{BagciTirn2009}. Then, the derived MaxEnt probability distribution is of ordinary exponential type and it results simultaneously and consistently the correct interval of concavity for the deformation parameter associated with the $S_{\text{\tiny{R}}}$ \cite{BagciTirn2009}. It is worth noticing that the same $q$-range is obtained through GMC-method. It is indeed surprising that the maximization of R\'enyi entropy so far was carried out  through linear averaged constraints. However, one can even obtain inverse power law stationary distributions from the maximization of BG entropy with arbitrary constraints if one is willing to do arbitrary physics. For serious physicists though, this is not an option.\\

Further, I would like to present some more evidence about the correctness of GMC-approach following a different mathematical route.
The standard method in computing MaxEnt distributions is by applying JF to a functional including the entropy measure and the suitably averaged constraints.
In case of BG-statistics, the results obtained from the former formalism coincide with the ones derived within thermostatistical theory.
However, once we apply JF on generalized entropic structures, there are no reference results to compare with and assure their correctness.
Indeed, the probabilities are mathematically dependent quantities through the normalization condition, thus the introduction of this dependence as a constant in JF assuming initially mathematical independence between the probabilities may lead generally to wrong results.
So, the idea is to check which probability distribution structure for a given entropy expression is compatible with the fundamental equilibrium relation of Thermodynamics, namely
%
\begin{equation}\label{eq:1}
dQ=dE+d\mc{W},
\end{equation}
%
where $Q$, $E$ and $\mc{W}$ is the heat flow, the internal energy and the produced work of the physical system under consideration, without computing derivatives. If we can determine the aforementioned structure, we can conclude that even the MaxEnt distribution must be of the same form. If we choose the initial values equal to zero, the above equation is written as
%
\begin{equation}\label{eq:2}
Q=E+\mc{W}.
\end{equation}
%
The statistical representation of this equation of ordinary statistics is given as
%
\begin{equation}\label{BasicRelation}
\av{\frac{1}{\beta}\ln(1/P_i)}_{\mt{lin}}=\av{\varepsilon_i}_{\mt{lin}}+\av{\mc{W}}_{\mt{lin}}
            \quad\Longrightarrow\quad
\av{\frac{1}{\beta}\ln(1/P_i)-\varepsilon_i-\mc{W}}_{\mt{lin}}=0\,,
\end{equation}
%
From the right hand relation of Eq. \eqref{BasicRelation}, it becomes obvious that in the ordinary case the probabilities which are compatible with the BG-entropic structure and the thermodynamic relation \eqref{eq:1} are of the following structure
%
\begin{equation}\label{BG-prob}
1/P_i   =\ln^{-1}\rbr{\beta\varepsilon_i+\beta\,\mc{W}}
        =\exp\Big(\beta\rbr{\varepsilon_i+\mc{W}}\Big)\,.
\end{equation}
%
This is the probability distribution obtained within the ordinary theory of Thermodynamics. Indeed, if we replace the above probability distribution in the probabilistic BG-heat expression $Q=\av{\frac{1}{\beta}\ln(1/P_i)}_{\mt{lin}}$, we obtain again the fundamental equilibrium thermodynamic relation in Eq. \eqref{eq:2}. Comparing Eqs. \eqref{BG-prob} and \eqref{BG-ConfFun} we see that the statistical quantity $W$ takes the form $W=\exp\rbr{\frac{1}{N}
(\varepsilon_i+\mc{W})}$ in terms of physical quantities.\\

Considering the heat expression $Q$, we observe two basic possibilities to generalize it, either we generalize the
\emph{uncertainty} $\ln(1/P_i)\ra\ln_\xi(1/P_i)$ or the averaging procedure  $\av{\cdots}_{\mt{lin}}\ra\av{\cdots}_{\sigma}$ (or both of them), where $\xi=\{\xi_i\}_{i=1,\ldots,u}$ and $\sigma=\{\sigma_i\}_{i=1,\ldots,u}$ are two deformation parameter sets. For certain values of these parameters $\xi\ra\xi_0$ and $\sigma\ra\sigma_0$ the generalized expression recovers the ordinary one in Eq. \eqref{BasicRelation}.
In the first case of generalized \emph{uncertainty} (trace-form entropies) Eq. \eqref{BasicRelation} tends to
%
\begin{equation}\label{BasicRelation-1}
\av{\frac{1}{\beta}\ln_\xi(1/P_i)}_{\mt{lin}}=\av{\varepsilon_{i}}_{\mt{lin}}+\av{\mc{W}}_{\mt{lin}}
\quad\Longrightarrow\quad
\av{\frac{1}{\beta}\ln_\xi(1/P_i)-\varepsilon_i-\mc{W}}_{\mt{lin}}=0
\end{equation}
%
and the obtained probability distribution is given by
%
\begin{equation}\label{eq:3}
1/P_{i}=\exp_\xi\Big(\beta\rbr{\varepsilon_{i}+\mc{W}}\Big).
\end{equation}
%
The distribution structure in Eq. \eqref{eq:3} is the one obtained from GMC$_1$ in Ref. \cite{Oik2007} for Tsallis entropy. We also see here that all trace-form entropies are compatible with linear averaged physical quantities. $P_i$ in Eq. \eqref{eq:3} preserves the extensivity property of the respective entropy, as explained in Ref. \cite{OikTirn2009}.\\

In the second case of the generalized nonlinear average (non-trace-form entropies) Eq. \eqref{BasicRelation} tends to
%
\begin{equation}\label{BasicRelation-2}
\av{\frac{1}{\beta}\ln(1/P_i)}_{\sigma}=\av{\varepsilon_{i}}_{\sigma}+\av{\mc{W}}_{\sigma},
\end{equation}
%
with $\av{\cdots}_{\sigma\ra\sigma_0}=\av{\cdots}_{\mt{lin}}$. The determination of the compatible probability distribution cannot be done in the general case, thus each non-trace form entropy should be studied separately. For the R\'enyi entropy, $\sigma\equiv\exp$, Eq. \eqref{BasicRelation-2}
takes the form
%
\begin{equation}
\av{\frac{1}{\beta}\ln(1/P_i)}_{\mt{exp}}=\av{\varepsilon_i}_{\mt{exp}}+\av{\mc{W}}_{\mt{exp}}
\quad\Longrightarrow\quad
\av{P_i^{\frac{1}{\beta}(q-1)}-\exp\Big((1-q)(\varepsilon_i+\mc{W})\Big)}_{\mt{lin}}=0\,.
\end{equation}
%
Then, the compatible probability distribution is of the form
%
\begin{equation}\label{eq:4}
1/P_i   =\exp\Big(\beta\rbr{\varepsilon_i+\mc{W}}\Big)\,.
\end{equation}
%
As can be seen, this is exactly the distribution one obtains with the GMC-approach in Ref. \cite{Oik2007}. Herewith, we complete the proof of the correctness of the results presented in Ref. \cite{Oik2007}, which are also confirmed by the general and consistent (contrary to the statement by Parvan who considers this maximization as a special case) maximization of R\'enyi entropy given in Ref. \cite{BagciTirn2009}. A crucial issue in the Eqs. \eqref{BasicRelation}-\eqref{eq:4} is whether the quantities $\varepsilon_i$ and $\mc{W}$ are the same in all ensembles. However, the answer to this question is not within the scope of this comment and deserves to be examined elsewhere in detail.\\

The author would like to thank U. Tirnakli for bringing Ref. \cite{Parvan2009} to my attention. G. B. Bagci and U. Tirnakli are acknowledged for carefully reading this comment and fruitful discussions.


\begin{references}

\bibitem{Parvan2009} A. S. Parvan, \textit{Critique of multinomial coefficients method for evaluating Tsallis and
                                 R\'enyi entropies},
                            Physica A, (2010) in press.


\bibitem{Oik2007} Th. Oikonomou, \textit{Tsallis R\'enyi and nonextensive Gaussian entropy derived from the respective
                            multinomial coefficients}
                            Physica A \textbf{386} (2007) 119.

\bibitem{Jaynes1965} E. T. Jaynes, \textit{Gibbs vs. Boltzmann entropies},
                            American J. Phys. \textbf{33}(No.5) (1965) 391.


\bibitem{Suyari2006} H. Suyari, \textit{Mathematical structure derived from the $q$-multinomial coefficient in Tsallis
                            statistics},
                            Physica A \textbf{368} (2006) 63.

\bibitem{Abe-Sisto} S. Abe, \textit{Remark on the escort distribution representation of nonextensive statistical
                            mechanics},
                            Phys. Lett. A \textbf{275} (2004) 223; R. P. Di Sisto, S. Martinez, R. B. Orellana,
                            A. R. Plastino, A. Plastino, \textit{General thermostatistical formalisms, invariance
                            under uniform spectrum translations, and Tsallis q-additivity},
                            Physica A \textbf{265} (1999) 590.

\bibitem{OikTirn2009} Th. Oikonomou \& U. Tirnakli, \textit{Generalized entropic structures and non-generality of
                            Jaynes' Formalism},
                            Chaos, Solitons and Fractals \textbf{42} (2009) 3027.


\bibitem{ParvanBiro2009} A. S. Parvan \& T. S. Bir\'o, \textit{R\'enyi statistics in equilibrium statistical
                            mechanics},
                            Phys. Lett. A \textbf{374} (2010) 1951.

\bibitem{BagciTirn2009} G. Baris Bagci \& Ugur Tirnakli, \textit{On the way towards a generalized entropy maximization
                            procedure},
                            Phys. Lett. A \textbf{373} (2009) 3230.




\end{references}
\end{document}